# The Rank-Size Scaling Law and Entropy-Maximizing Principle

Yanguang Chen

(Department of Geography, College of Urban and Environmental Sciences, Peking University, Beijing 100871, P.R. China. Email: chenyg@pku.edu.cn )

**Abstract**: The rank-size regularity known as Zipf's law is one of scaling laws and frequently observed within the natural living world and in social institutions. Many scientists tried to derive the rank-size scaling relation by entropy-maximizing methods, but the problem failed to be resolved thoroughly. By introducing a pivotal constraint condition, I present here a set of new derivations based on the self-similar hierarchy of cities. First, I derive a pair of exponent laws by postulating local entropy maximizing. From the two exponential laws follows a general hierarchical scaling law, which implies general Zipf's law. Second, I derive a special hierarchical scaling law with exponent equal to 1 by postulating global entropy maximizing, and this implies the strong form of Zipf's law. The rank-size scaling law proved to be one of the special cases of the hierarchical law, and the derivation suggests a certain scaling range with the first or last data point as an outlier. The entropy maximization of social systems differs from the notion of entropy increase in thermodynamics. For urban systems, entropy maximizing suggests the best equilibrium state of equity for parts/individuals and efficiency for the whole.

# 1 Introduction

If a country or a region is large enough to encompass a great many cities, these cities usually follow the well-known Zipf's law (Zipf, 1949). Zipf's law is associated with the rank-size rule, and the former does not differ from the latter in practice. Formally, the rank-size scaling law can be expressed as

$$P_k = P_1 k^{-q}, \tag{1}$$

where $k$ denotes the rank by size of cities in the set ($k$=1, 2, 3, …), $P_k$ refers to the population size of the $k$th city, $P_1$ to the population of the largest city, and $q$, the scaling exponent of the rank-size distribution. Empirically, $q$ values always approach to 1 (Basu and Bandyapadhyay, 2009; Gabaix, 1999a; Gabaix, 1999b; Ioannides and Overman, 2003; Jiang and Yao, 2010; Krugman, 1996; Saichev *et al*, 2011; Zipf, 1949). The rank-size regularity is associated with fractals, and the power exponent indicates the fractal dimension of city-size distributions (Batty, 2006; Batty and Longley, 1994; Chen and Zhou, 2003; Frankhauser, 1998; Mandelbrot, 1983; Salingaros and West, 1999; Semboloni, 2008).

In theory, if we define a self-similar hierarchy of $M$ levels, with the $f_1$=1 city in the first level, $f_2$=$r_f$ cities in the second level, $f_3$=$r_f^2$ cities in the third level, and so on, the rank-size law, equation (1), can be decomposed as a pair of exponential laws in the forms

$$f_m = f_1 r_f^{m-1}, \qquad (2)$$

$$P_m = P_1 r_p^{1-m}, \qquad (3)$$

where $m$ denotes the level order of cities in the hierarchy ($m$=1, 2, 3, …, $M$), $f_m$ refers to the number of cities in the $m$th level, $P_m$, to the average size of the $f_m$ cities, $P_1$ to the average population size of the top-class cities ($f_1$=1), $r_f$ to the city *number ratio* ($r_f$=$f_{m+1}/f_m$), and $r_p$, city *size ratio* ($r_p$=$P_m/P_{m+1}$). Equation (2) represents the number law, and equation (3), the size law of cities. If $r_f$=$r_p$=2, equations (2) and (3) are equivalent to the $2^n$ rule of Davis (1978). Therefore, the pair of exponential functions represents the generalized $2^n$ rule of cities (Chen and Zhou, 2003).

As stated above, the self-similar hierarchy is based on a top-down order, i.e., from the largest to the smallest. Due to the mirror symmetry of exponential models, we can describe the hierarchy in the inverse order equivalently, that is, from the smallest to the largest (Chen, 2009). Based on the bottom-up order, the number law, equation (2), can be re-expressed in an equivalent form

$$f_m = f_M r_f^{1-m}, \qquad (4)$$

where $m$ denotes the order from the smallest to the largest cities, $f_M$=$f_1$' is the city number of the bottom level in the top-down order hierarchy, and $f_1$', the city number in the first level of bottom-up order hierarchy. Correspondingly, the size law can be rewritten as $P_m$=$P_M r_p^{m-1}$, in which $P_M$ indicates the average size of the smallest cities in the first kind of hierarchy. By the generalized $2^n$ principle, the rank-size scaling law can be reconstructed as a hierarchical scaling law (Chen,

2010). Obviously, from equations (2) and (3) follows a scaling relation in the form

$$f_m = \eta P_m^{-D}, \qquad (5)$$

where $\eta = f_1 P_1^D$ denotes a proportionality coefficient, and $D=1/q$ proved to be a fractal dimension of city rank-size distribution (Chen and Zhou, 2009).

Zipf's law is in fact one of the scaling law in nature and society (Bettencourt *et al*, 2007). However, for a long time, the rank-size rule is not derivable from the general principle so that no convincing physical and economic explanation can be provided for existence of the scaling relation and exponent value (Córdoba, 2008; Johnston *et al*, 1994; Vining, 1977). In order to bring to light the underlying rationale of the rank-size scaling of cities, scientists tried to derive it by using some approaches, say, the entropy-maximizing methods. It was shown that Zipf's law can be related to the maximum entropy models (Mora *et al*, 2010). Curry (1964) once made a derivation of the rank-size rule by the idea from entropy maximization. However, his demonstration has three bugs. First, he actually derived an exponential model associated with equation (4) rather than a power law, equation (1). Second, he let $f_m=m$, that is, city number in each class is confused with the order of the class. Third, one of the constraint conditions was set as $f_m P_m = const$, and this does not accord with the reality. Anastassiadis (1986) attempted to derive the equation (1) directly, but too many assumptions were made that the mathematical process is too complicated to be understood. Chen and Liu (2002) made another derivation with the entropy-maximizing principle based on Curry's work and the self-similar hierarchy, and they deduced equation (3) and (4) implying equations (2) and (5). However, the second and third problems of Curry's assumptions failed to be resolved so that the mathematical process and physical explanation are not yet very convincing.

In fact, urban evolution falls into two major, sometimes contradictory, processes: city number increase and city size growth (Steindl, 1968; Vining, 1977). The former indicates what is called *external complexity* associated with frequency distribution scaling, and the latter, *internal complexity* associated with size distribution scaling. The concepts of external and internal complexity came from biology (Barrow, 1995). In this paper, I will present a new derivation of the rank-size scaling law of city by the method of entropy-maximizing, and then propose a new explanation for Zipf's law. First, assuming entropy-maximizing of city frequency distribution, I derive equation (4), which is equivalent to equation (2); Second, assuming entropy-maximizing of city size distribution, I derive equation (3), and from equations (2) and (3) follows equation (5), which suggests equation (1); Third, assuming entropy-maximizing of both city frequency and size distributions, I derive the scaling exponent *q*=1. Two empirical analyses will be made to lend support to the theoretical results.

## 2 Models and derivations

### 2.1 Derivation of the number law

Suppose there is a region **R** with *n* cities and a total urban population of *N* inside. A basic assumption is made as follows: the probability of an urban resident living in different cities is equal to one another (Curry, 1964). By the average population sizes in different groups, we can classify the cities into *M* levels in bottom-up order. If the city number in the *m*th level is $f_m$ and the mean size of the $f_m$ cities is $P_m$, the state number of the *n* city distribution in different classes, *W*, can be expressed as a problem of ordered partition of the city set. In fact, an ordered partition of "type $f_1+f_2+\ldots+f_M$" is one in which the *m*th part has $f_m$ members, for *m*=1, 2, ..., *M*. The number of such partitions is given by the multinomial coefficient

$$W(f) = \binom{n}{f_1, f_2, \cdots, f_m, \cdots, f_M} = n! \bigg/ \prod_{m=1}^{M} f_m!, \tag{6}$$

where *m*=1, 2, …, *M* denotes the order of city classes. Thus the information entropy of frequency distribution of cities is

$$H_f = \ln W(f) = \ln n! - \sum_{m=1}^{M} \ln f_m!, \tag{7}$$

where $H_f$ refers to the information entropy. Suppose that entropy maximization is the objective of urban evolution. A nonlinear "programming" model can be built as follows

$$\text{Max} \quad H_f = \ln W(f), \tag{8}$$

$$\text{S.t.} \quad \sum_m f_m = n, \tag{9}$$

$$\sum_m m f_m = n^2 / f_M. \tag{10}$$

This is a typical optimization problem. The first constraint condition, equation (9), is easy to understand, and it indicates that the city number in the region is certain at given time. The second constraint condition, equation (10), will be specially explained in Section 3. In fact, equation (10) implies equation (9), and the latter will be demonstrated to be excrescent.

In order to find the conditional extremum, i.e., the maxima or minima of a function subject to constraints, we can construct a Lagrange function such as

$$L(f) = \ln n! - \sum_m \ln f_m! + \lambda_1(n - \sum_m f_m) + \lambda_2(n^2/f_M - \sum_m m f_m), \qquad (11)$$

where $\lambda_1$ and $\lambda_2$ denote Lagrange multipliers. The Lagrange multipliers can provide a strategy of yielding a necessary condition for optimality in constrained problems (Bertsekas, 1999). If a number $x$ is large enough, then, according to Stirling's formula $x! = \sqrt{2\pi} x^{x+1/2} e^{-x}$, we will have an approximate expression $d\ln x!/dx \approx \ln x$. So, if $n$ and $f_m$ are large enough in theory, then $d\ln n!/dn \approx \ln n$, $d\ln f_m!/df_m \approx \ln f_m$. Considering Lagrangian condition of extreme value $\partial L(f)/\partial f_m = 0$, we can find that $\lambda_1 \approx 0$, $\lambda_2 = \ln(r_f)$, and have

$$f_m = f_0 e^{-\omega m} = f_1' r_f^{1-m}, \qquad (12)$$

where $f_0 = f_M r_f$, $f_1' = f_M$, $\omega = \lambda_2 = \ln(r_f)$. This is just the equivalent expression of the number law, i.e., equation (1). The first derivation is complete.

## 2.2 Derivation of the size law

The basic assumptions given is subsection 2.1 hold on, and let $P$ represent the summation of the average urban population in different classes, i.e., $P = \sum P_m$. The state number of the average population size distribution in the hierarchy, $W(P)$, can be expressed as an ordered partition and defined by

$$W(P) = \begin{pmatrix} P \\ P_1 \ P_2 \ \cdots \ P_M \end{pmatrix} = P! / \prod_{m=1}^{M} P_m!. \qquad (13)$$

Accordingly, the size distribution information entropy ($H_P$) function is

$$H_P = \ln W(P) = \ln P! - \sum_{m=1}^{M} P_m. \qquad (14)$$

Suppose that the information entropy of city development approaches maximization, a nonlinear programming model can be made in the form

$$\text{Max} \quad H_P = \ln W(P), \qquad (15)$$

$$\text{S.t.} \quad \sum_m P_m = P, \tag{16}$$

$$\sum_m mP_m = P^2 / P_1. \tag{17}$$

To find the optimum solution to the optimization problem, we can set a Lagrange function

$$L(P) = \ln P! - \sum_m \ln P_m! + \lambda_3 (P - \sum_m P_m) + \lambda_4 (P^2 / P_1 - \sum_m mP_m), \tag{18}$$

where $\lambda_3$ and $\lambda_4$ represent Lagrange multipliers. According to Stirling's approximation, if $P$ and $P_m$ are large enough, then $d\ln P!/dP \approx \ln P$, $d\ln P_m!/dP_m \approx \ln P_m$. Considering the condition of extreme value $\partial L(f)/\partial f_m = 0$, we find that $\lambda_3 \approx 0$, $\lambda_4 = \ln(r_p)$, and thus

$$P_m = P_0 e^{-\varphi m} = P_1 r_p^{1-m}, \tag{19}$$

where $P_0 = P_1 r_p$, $\varphi = \lambda_4 = \ln(r_p)$. This is just the size law, i.e., equation (2). The second derivation is concluded.

From equation (12) and equation (19) follows equation (5), which thus implies equation (1), the general form of Zipf's law. The derivation of the city population size law, equation (2), can be generalized to urban area or urban land use size law, thus, we can derive the allometric scaling relation between urban area and population (Chen, 2009).

## 2.3 Derivation of the standard scaling exponent

If the whole hierarchy of cities conforms to the principle of entropy maximization, we can demonstrate that the scaling exponent approaches 1, i.e., $D \to 1$. In terms of the postulate *ut supra*, the city number in each class is

$$f_m P_m = S_m, \tag{21}$$

The state number of total urban population in the hierarchy can be given by

$$W(S_m) = \binom{N}{S_1 \; S_2 \; \cdots \; S_M} = N! / \prod_{m=1}^{M} S_m!. \tag{22}$$

Thus the information entropy function is

$$H_S = \ln N! - \sum_{m=1}^{M} \ln S_m!. \tag{23}$$

Suppose finding the maximum of the entropy function $H_S = \ln W(S)$ subject to constraint $\sum S_m = N$. This means that entropy is maximized on condition that the summation of city population of different classes equals *N*. If *M* is limited, a Lagrange function can be defined by

$$L(S) = \ln N! - \sum_m \ln S_m! + \lambda(\sum_m S_m - N). \tag{24}$$

According to the condition of extreme value, derivative of *L*(*S*) with respect to *S* yields

$$S_m = e^\lambda = \eta = f_m P_m. \tag{25}$$

Thus we have $\lambda=\ln(P_1)=\ln(N/M)$. Rearranging equation (25) givens a special inverse power function

$$f_m = \eta P_m^{-1}. \tag{26}$$

Comparing equation (26) with equation (3) suggests that

$$D = \frac{\ln(f_{m+1}/f_m)}{\ln(P_m/P_{m+1})} = \frac{\ln r_f}{\ln r_p} = \frac{\omega}{\varphi} = 1. \tag{27}$$

The proof is over and the conclusion can be drawn that the fractal parameter of the hierarchical scaling approaches 1 due to maximizing information entropy of urban systems. In this instance, $q=1/D=1$, and we have the strong form of the Zipf distribution (Batty, 2006).

## 3 Discussion

### 3.1 Constraint equations and scaling range

According to information theory, there are three probability density distribution functions which are associated with the principle of entropy maximization. That is, uniform distribution (a.k.a. rectangular distribution), exponential distribution (a.k.a. negative exponential distribution), and normal distribution (a.k.a. Gaussian distribution) (Chen, 2009; Silviu, 1977; Reze, 1961). If the variable *x* has clear upper limit *a* and lower limit *b*, the probability density based on entropy maximization satisfies the uniform distribution; If the variable *x* has clear lower limit *a*=0, but no upper limit (*b* goes to infinity), the entropy-maximization-based probability density meets the exponential distribution; If the variable *x* has neither lower limit nor upper limit (both *a* and *b* go to infinity), the probability density based on entropy maximization takes on the normal distribution (Table 1). In Subsections 2.1 and 2.2, we derived two exponential distribution functions; in Subsection 2.3, we in fact derive a uniform distribution. Exponential distribution and uniform distribution can always be derived by entropy-maximizing methods. The approach is clear, and the derivation is simple, but the difficulty and key of a study depend on the assumptions and construction of constraint equations.

Table 1 Three probability distribution functions based on entropy-maximization principle

| Interval | Probability density function | Characteristic scale | Maximum entropy |
|---|---|---|---|
| $a \le x \le b$ | $f(x) = \dfrac{1}{b-a}$ | Range: $b$-$a$ | $\ln(b-a)$ |
| $0 \le x < \infty$ | $f(x) = \dfrac{1}{\mu}\exp(-\dfrac{x}{\mu})$ | Mean value: $\mu$ | $\ln(\mu e)$ |
| $-\infty < x < \infty$ | $f(x) = \dfrac{1}{\sqrt{2\pi}\sigma}\exp(-\dfrac{x^2}{2\sigma^2})$ | Standard error: $\sigma$ | $\ln(\sqrt{2\pi}\sigma e)$ |

**Note**: This table is summarized by reference to Reza (1961, page 268) and Silviu (1977, page 298). As for the parameters, $\mu$ refers to average value, $\sigma$ to standard error, and $e$, the base of the natural system of logarithms, having a numerical value of approximately 2.7183.

One of the problems in the previous derivation lies in that the main constraint is beyond belief. A key of the originalities in this paper rests with finding the constraint equations. Suppose that the urban population in region **R** is infinite. The total urban population $N$ sometimes approaches infinity as a limit, thus $\sum f_m P_m = N$ does not always converge. However, we can demonstrate that $\sum f_1 m r_f^{1-m} = n^2/f_1$ and $\sum P_1 m r_p^{1-m} = P^2/P_1$ will converge to a certainty. Consider a geometric series. For $x<1$, the summation is

$$s = \sum_m x^{m-1} = \frac{1}{1-x}, \qquad (28)$$

It is easy to prove the following relation

$$T = \lim_{m \to \infty} \sum_m m x^{m-1} = s^2 = \frac{1}{(1-x)^2}. \qquad (29)$$

In fact, the first derivative of the geometric progression is as follows

$$\begin{aligned}
T &= 1 + 2x + 3x^2 + \cdots + mx^{m-1} + \cdots \\
&= (1 + x + x^2 + \cdots + x^{m-1} + \cdots) + (x + x^2 + x^3 + \cdots + x^{m-1} + \cdots) \\
&\quad + (x^2 + x^3 + x^4 + \cdots + x^{m-1} + \cdots) + \cdots + (x^{m-1} + x^m + x^{m+1} + \cdots) + \cdots. \\
&= s + xs + x^2 s + \cdots + x^{m-1} s + \cdots = s(1 + x + x^2 + \cdots + x^{m-1} + \cdots) \\
&= s^2
\end{aligned}$$

This suggests that if a geometric series converges, the first derivative of the geometric series will converge all to nothing. The inference can be verified by a simple mathematical experiment. This implies that, for the objective of entropy maximization, if

$$\sum_m kx^{m-1} = ks = h \tag{30}$$

is one of constraint conditions, we will have an alternative such as

$$\sum_m kmx^{m-1} = ks^2 = h^2/k, \tag{31}$$

where $h$ and $k$ are constants. Equations (9) and (16) are actually redundant so that $\lambda_1$ and $\lambda_3$ equal zero.

The derivation of the pair of exponential laws actually implies the derivation of scaling range of the hierarchical power law. According to Stirling's approximate formula, only if $f_m$ is large enough, the number law, equation (4) and thus equation (2), will be derivable, and only if $P_m$ is large enough, the size law, equation (3), will be derivable. If and only if equation (2) and equation (3) are derivable, equation (5) and thus equation (1) can be derived by the entropy-maximizing methods. This suggests that the largest city or even more than one top city does not always fall into the valid scaling range. What is more, the cities in the "ground floor" (bottom level) also make an exception in the empirical scaling analysis. Therefore, generally speaking, only the middle data points form a straight line segment on a rank-size or number-size log-log plot (double logarithmic plot). The slope of the line segment within the scaling range gives the scaling exponent. The first or the last data point usually makes an outlier of a scaling relation.

## 3.2 Duality of city development

The derivation of the rank-size rule and the scaling exponent value suggests that the process of urban evolution is indeed related to the principle of entropy-maximizing. The entropy maximization of human systems differs from the concept of entropy increase in thermodynamics (Anastassiadis, 1986; Bussière and Snickars, 1970; Chen, 2009; Curry, 1964; Wilson, 2000). Where city development is concerned, the entropy-maximizing implies the unity of opposites between the equity for parts or individuals and efficiency for the whole. For example, based on equation (23), a dual nonlinear programming model can be built to show the relation between individuals/parts and the whole. Maximizing the entropy subjective to certain urban population yields the primal problem such as

$$\begin{aligned} \text{Max} \quad & H_S = \ln N! - \sum_m \ln S_m! \\ \text{s.t.} \quad & \sum_m S_m = N \end{aligned} \quad . \tag{32}$$

The dual problem is to minimize the urban population subjective to determinate entropy, that is

$$\begin{aligned} \text{Min} \quad & N = \sum_m S_m \\ \text{s.t.} \quad & \ln N! - \sum_m \ln S_m! = H_S \end{aligned} \quad . \tag{33}$$

The primal problem and the dual problem share the common resolution, that is, the city number of each class is in inverse proportion to the average population size of the corresponding class, which can be formulated as equation (26). The effective strategy for finding the maxima or minima of an optimization function subject to constraints is the method of Lagrange multipliers.

The nonlinear programming problems are ones of optimization problems. In terms of the primal model, entropy maximizing suggests equity, equity means equality and justice, but does not imply average. The different classes of the self-similar hierarchy represent different "energy levels" of urban population distribution (Figure 1). Geographical space is not a kind of homogeneous space. The energy level of a city accords with its geographical conditions. The better the geographical condition is, the larger the city will become, and thus the higher the energy level of the city will be. The higher the energy level is, the fewer the city number will get. For the standard case, the product of city number and average city size is a constant, that is

$$f_m P_m = \eta \to P_1 \quad . \tag{34}$$

On the other hand, in light of the dual model, city-number minimizing suggests efficiency. Nature pursues frugality and economization. If 100 cities are enough to accommodate all the urban population in a region, does not build 101 cities. This suggests the principle of least effort of human behavior (Ferrer i Cancho and Solé, 2003; Zipf, 1949), which seems to be associated with the principle of least action in physics.

The self-similar hierarchy of cities is not the best model showing the relation between equity and efficiency in terms of entropy-maximization. The best case may be Wilson's spatial interaction model (SIM) on traffic flows (Wilson, 1971; Wilson, 2000). According to SIM, we can maximize entropy of traffic flow distribution subjective to given transport costs, or minimize the transport costs subjective to certain entropy. Without question, transport cost minimizing indicates traffic efficiency (least effort or least action). On the other hand, entropy-maximizing suggests that the product of traffic flow quantity and transport cost is a constant. This is a kind of condition equality: the better the transport condition is, the more the traffic flow quantity will be.

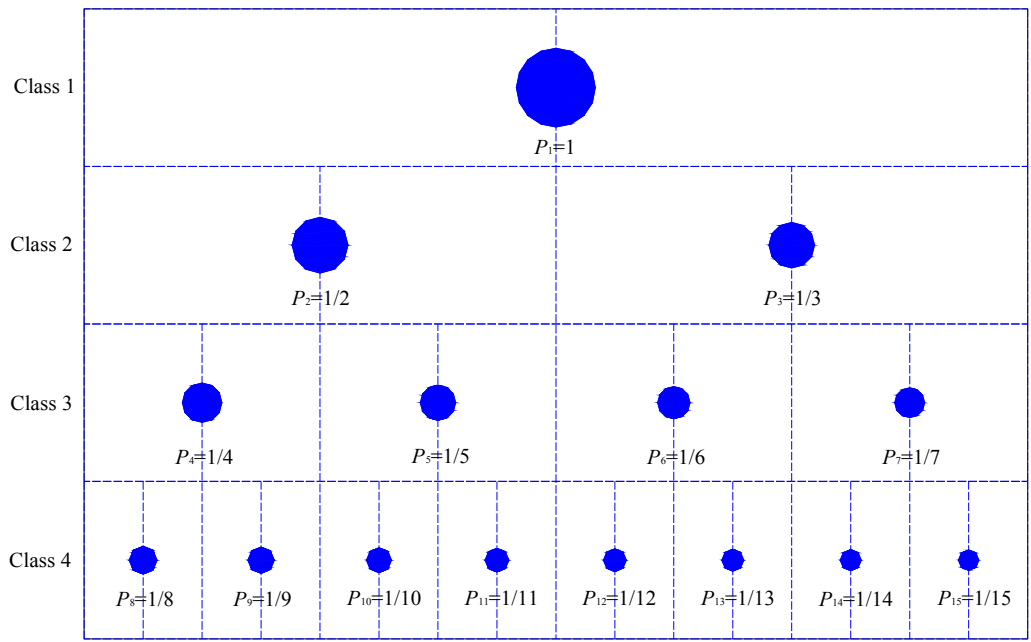

**Figure 1** The "energy level" of cities based on the hierarchy with cascade structure

## 4 Conclusions

The rank-size scaling regularity is a special case of the hierarchical scaling law of cites, and the hierarchical scaling relation can be derived by the entropy maximizing methods. First, a pair of exponential laws can be derived from the postulates of local entropy maximization. Assuming the frequency distribution entropy maximizing, we can derive the number law of urban hierarchies; assuming the size entropy maximizing, we can derive the size law of hierarchies of cities. From the number law and size law follows the hierarchical scaling law, which implies the general rank-size scaling law. Second, the special scaling exponent, 1, for the strong form of Zipf's law, can be derived from the postulate of global entropy maximization. The fractal dimension $D=1$ suggests the equilibrium of frequency distribution entropy maximizing (external complexity) and size distribution entropy maximizing (internal complexity). Third, the derivation processes of the number law and the size law suggest a scaling range on the rank/number-size log-log plot for empirical analysis. Fourth, city development is a process of unity of opposites. Urban evolution seems to struggle between order and chaos, finitude and infinitude (say, city number is limited or limitless), simplicity and complexity (say, exponential distribution and power-law distribution), internal complexity and external complexity (say, city number increase and size growth), and so on. Fifth, entropy maximizing indicates complication and optimization of city development. Where urban evolution is concerned, entropy maximization suggests the equity for parts or individuals and efficiency for the whole harmonizes to the best state.

## 5 Materials and methods

All the theoretical results can be directly or indirectly supported by the empirical observations in the real world (Basu and Bandyapadhyay, 2009; Chen and Zhou, 2003; Gabaix, 1999a; Gabaix, 1999b; Gangopadhyay and Basu, 2009; Ioannides and Overman, 2003; Krugman, 1996; Jiang and Jia, 2011; Moura and Ribeiro, 2006; Peng, 2010). The rank-size scaling is actually a special case of the hierarchical scaling, and the Zipf distribution is a special case of hierarchical scaling relations. Zipf's law can be regarded as one of the signatures of the self-similar hierarchical structure. If the fractal parameter $D=1$, equation (5) will change to equation (26). By l'Hôpital's rule (a.k.a. Bernoulli's rule), let $r_f=r_p=1$, we can derive the strong form of Zipf's law from equation (26) such as

$$P_k = P_1 k^{-1}. \qquad (35)$$

If $r_f=r_p=2$, we will have Davis' $2^n$ rule (Chen and Zhou, 2003; Davis, 1978). In fact, we can let $r_f=r_p=3, 4, 5, 6\ldots$, and the hierarchical scaling relation associated with equation (35) will not break down. This can be verified by simple mathematical experiments with MS Excel or other mathematical software. Therefore, the derivation of the hierarchical scaling laws suggests the derivation of the rank-size scaling law by the entropy maximizing methods. Now, I will employ several empirical cases to support the following judgment: if elements in a real system, say, system of cities, follow Zipf's law or the rank-size rule, the observational data can be fitted to the hierarchical scaling relation and *vice versa*.

The two cases are empirically consistent with the rank-size distribution (Figure 2). The first case is the system of the cities of the United States of America. According to the US census, there are 452 cities with population size over 50,000 by urbanized area in 2000. The data are available from internet (see the note below Table 2). A least square computation gives the following rank-size relation

$$P_k = 52516701.468 k^{-1.125} \ .$$

The correlation coefficient square is about $R^2$=0.9893, and the fractal dimension of city-size distribution is estimated as $D=1/q \approx 1/1.125 \approx 0.889$. The second case is the systems of the Indian cities. The statistical dataset of the top 300 cities of India in 2000 are available from internet (see also the note below Table 2). A least square calculation yields the following result

$$P_k = 17702906.650 k^{-0.842} \ .$$

The goodness of fit is around $R^2$=0.9944, and the fractal parameter of city-size distribution is estimated as $D \approx 1/0.842 \approx 1.188$.

Table 2 The self-similar hierarchies of the US cities and India cities in 2000

| Order | America's cities | | | India's cities | | |
|---|---|---|---|---|---|---|
| $m$ | Number $f_m$ | Total $S_m$ | Size $P_m$ | Number $f_m$ | Total $S_m$ | Size $P_m$ |
| 1 | 1 | 17799861 | 17799861.000 | 1 | 12622500 | 12622500.000 |
| 2 | 2 | 20097391 | 10048695.500 | 2 | 15253700 | 7626850.000 |
| 3 | 4 | 18246258 | 4561564.500 | 4 | 16393400 | 4098350.000 |
| 4 | 8 | 26681941 | 3335242.625 | 8 | 18137700 | 2267212.500 |
| 5 | 16 | 27052740 | 1690796.250 | 16 | 19090200 | 1193137.500 |
| 6 | 32 | 26098069 | 815564.656 | 32 | 25357200 | 792412.500 |
| 7 | 64 | 22690390 | 354537.344 | 64 | 25985200 | 406018.750 |
| 8 | 128 | 19988240 | 156158.125 | 128 | 27509300 | 214916.406 |
| 9[*] | 197 | 13738825 | 69740.228 | 45 | 6888500 | 153077.778 |

Note: The original city dataset of America is available from: http://www.demographia.com/db-ua2000pop.htm ; The city dataset of India is available from: http://www.tageo.com/index-e-in-cities-IN.htm .* The last order is a lame-duck class of Davis (1978) due to absence of enough city data.

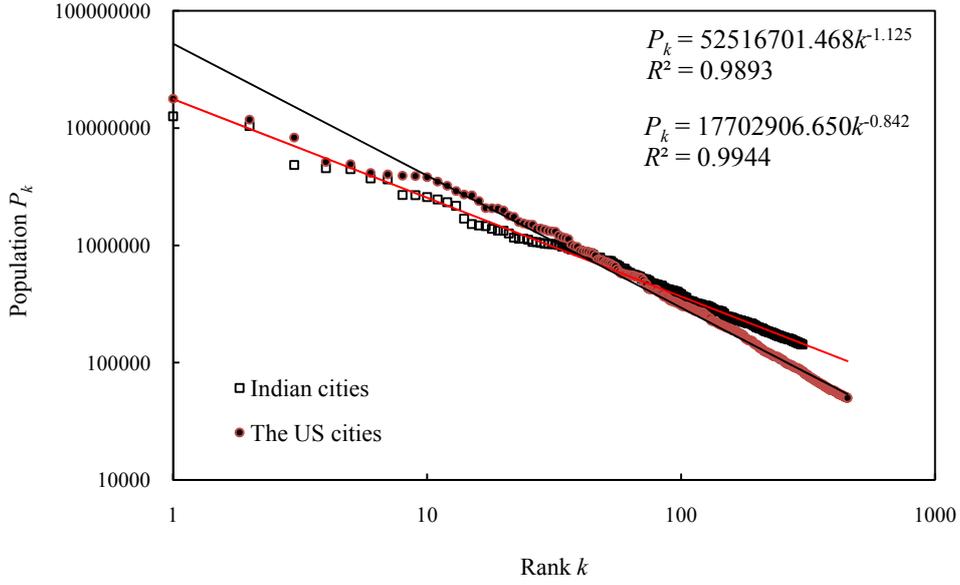

**Figure 2** The rank-size patterns of the US cities and Indian cities in 2000 (the solid dots indicate the 452 US cities, the hollow squares denote the 300 cities of India)

In light of Figure 1, defining a number ratio $r_f=2$ as an *ad hoc* value, we can construct a self-hierarchy for both the 452 US cities and the 300 Indian cities in 2000 (Table 2). The number of levels is $M=9$, thus the city number in the bottom order is expected to be $f_M=f_1 r_f^{m-1}=2^8=256$. However, owing to absence of enough city data or undergrowth of cities, the number of the cities in the ground floor is in fact $f_9=197<256$ for US and $f_9=45<256$ for India. In this instance, the last level is a "lame-duck class" termed by Davis (1978), so the corresponding data points are removed as outliers from the regression analysis (Figure 3). The hierarchical scaling relation of the US cities is

$$f_m = 36836707.378 P_m^{-1.035}.$$

The goodness of fit is about $R^2=0.9909$, and the fractal dimension is estimated as $D\approx1.035$. The size ratio is about $r_p=1.994$, which gives another fractal dimension estimation $D=\ln r_f/\ln r_p\approx1.004$. The hierarchical scaling equation of Indian cities is as below

$$f_m = 307810420.013 P_m^{-1.193}.$$

The goodness of fit is around $R^2=0.9986$, and the fractal dimension is about 1.193, close to 1.188. The size ratio is about $r_p=1.796$, which gives another fractal dimension estimation $D\approx1.184$.

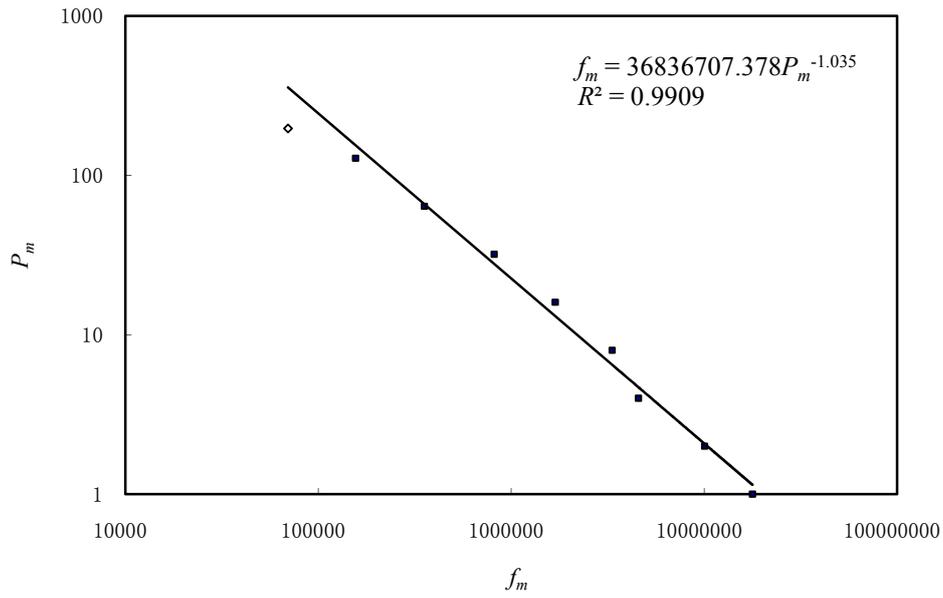

a. The US cities

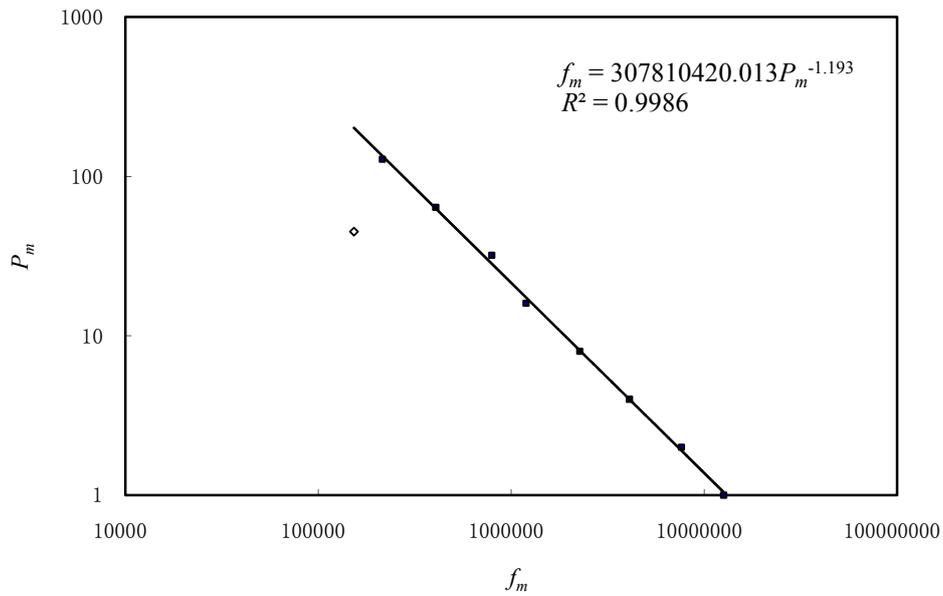

b. Indian cities

**Figure 3** The hierarchical scaling patterns of the US cities and Indian cities in 2000 (the solid squares denote the data points within the scaling ranges, and the diamond-shaped symbols indicate the outliers in the absence of enough cities in the bottom classes)

Both the US cities and Indian cities follow the rank-size scaling law and hierarchical scaling law at the same time. But there is subtle difference between the US cities and Indian cities. Where Indian cities are concerned, the exponent values from the hierarchical scaling analysis (1.193±0.018, 1.184) are very close to the result from the rank-size scaling analysis (1.188).

However, as far as the US cities are concerned, the hierarchical scaling exponent (1.035±0.041, 1.004) is not very close to the rank-size scaling exponent value (0.889±0.004). Tracing this difference to its source, we can find the trail of scaling break in the rank-size distribution of the US cities. For the 300 largest Indian cities, almost all the data points distribute along a single trend line on the double logarithmic plot. However, for the 452 top US cities, the data points actually distribute along two trend lines with different slopes. The large cities in the minority (about 32 cities) share one trend line (the slope is about $q$=0.8), while the medium-sized and small cities (about 420 cities) in the majority share another trend line (the slope is about $q$=1.2). On the whole, the slope is about $q$=1. This is consistent with the principle of entropy maximization. As shown above, when and only when the city number $f_m$ is large is enough, the number law can be derived, and thus the rank-size scaling law can be transformed into the hierarchical scaling relation.

As the scaling exponent values of the hierarchy of the US cities is near 1, we can also define $r_f$=3, 4, and 5. The hierarchical scaling exponent changes little with the number ratio. As for Indian cities, since the scaling exponent is not close to 1, we had better take $r_f$=2 rather than other values. Conclusions can be reached as follows. First, the hierarchical scaling indeed implies the rank-size scaling. The rank-size scaling can be derived from the hierarchical scaling, while the hierarchical scaling relation can be derived with the entropy maximizing method. Therefore, the rank-size scaling does come from the principle of entropy-maximization. Second, if we employ the hierarchical scaling analysis to estimate the scaling exponent of the rank-size distribution, we get the average value of the slopes of different scaling ranges. In the case of scaling break to some extent, the hierarchical scaling analysis has an advantage over the rank-size scaling analysis. Third, the scaling exponent can be used to judge whether or not the city size distribution is close to the optimum state. If and only if the scaling exponent approaches 1, the two kind of structural entropies reach the equilibrium, and the elements of the urban system become of harmony.

A question is that the scaling exponent estimation of city size distribution depends on the definition of cities. So far, the city has been a subjective notion in various countries (Jiang and Yao, 2010; Rozenfeld *et al*, 2008). For example, there are three basic concepts of cities in US: city proper (CP), urbanized area or urban agglomerations (UA), and metropolitan area (MA) (Davis, 1978; Rubenstein, 2007). Fit the population size data of the largest 600 US cities by CP in 1990 and 2000 (census) to equation (1) yields scaling exponents around $q$=3/4. If and only if we use the UA data to estimate the scaling exponent, the result is close to $q$=1. To estimate objective value of the scaling exponent, we need an objective city definition. Maybe a solution to this problem is to employ the "natural cities" defined by Jiang and his co-workers (Jiang and Jia, 2011; Jiang and Liu, 2011). The "natural city" is the most objective definition of cities which we can find at present.

**Acknowledgements:**

This research was sponsored by the National Natural Science Foundation of China (Grant No. 40771061. See: https://isis.nsfc.gov.cn/portal/index.asp). The support is gratefully acknowledged.

# Appendices

## A1 Detailed derivation of the exponential laws (an example)

Derivative of the Lagrange function, equation (11), with respect to $f_m$ yields

$$\frac{\partial L(f)}{\partial f_m} = \ln n - \ln f_m - \lambda_1 - \lambda_2 m. \tag{a1}$$

Considering the condition of the extremum, we have

$$f_m = n e^{-\lambda_1} e^{-\lambda_2 m} = n e^{-\lambda_1 - \lambda_2} e^{-\lambda_2 (m-1)}. \tag{a2}$$

Given $m=1$, it follows that

$$f_1' = n e^{-\lambda_1 - \lambda_2} = r_f^{M-1}; \tag{a3}$$

On the other hand, if $m=M$ as given, then

$$f_M' = n e^{-\lambda_1 - \lambda_2} e^{-\lambda_2 (M-1)} = f_1' e^{-\lambda_2 (M-1)} = 1, \tag{a4}$$

which implies

$$e^{-\lambda_2} = f_1'^{1/(1-M)} = 1/r_f. \tag{a5}$$

Thus we get

$$\lambda_2 = \frac{\ln f_1'}{M-1} = \frac{\ln r_f^{M-1}}{M-1} = \ln r_f. \tag{a6}$$

Substituting equation (a5) into equation (a3) yields

$$e^{-\lambda_1} = \frac{f_1}{n} e^{\lambda_2} = \frac{r_f f_1}{n} = \frac{r_f^M}{n} \approx 1. \tag{a7}$$

So equation (a2) can be rewritten as

$$f_m = n e^{-(\ln r_f)m} = f_1' e^{-\ln(r_f)(m-1)} = f_M r_f^{1-m}. \tag{a8}$$

According to the constraint equation (9), we have

$$\sum_m f_m = n e^{-\lambda_1 - \lambda_2} \sum_m e^{-\lambda_2(m-1)} = n e^{-\lambda_1 - \lambda_2} \left(\frac{1}{1-e^{-\lambda_2}}\right) = \frac{f_1}{1-e^{-\lambda_2}} = n; \tag{a9}$$

According to the constraint equation (10), we get

$$\sum_m mf_m = ne^{-\lambda_1-\lambda_2}\sum_m me^{-\lambda_2(m-1)} = ne^{-\lambda_1-\lambda_2}(\frac{1}{1-e^{-\lambda_2}})^2 = f_1(\frac{1}{1-e^{-\lambda_2}})^2 = \frac{n^2}{f_1}. \quad (a10)$$

Obviously equation (a10) implies equation (a9), and this suggests that the constraint equation (9) is not necessary for our derivation.

By analogy, we can derive equation (19) and its parameters' mathematical expression.

**A2 Four hierarchies of cities based on the concept of "natural city"**

Recently, Bin Jiang and his coworkers have proposed a concept of "natural city" and developed a novel approach to measuring objective city sizes based on street nodes or blocks and thus urban boundaries can be naturally defined (http://arxiv.org/find/all/). The street nodes are defined as street intersections and ends, while the naturally defined urban boundaries constitute what is called *natural cities*. The street nodes are significantly correlated with population of cities as well as city areal extents. The city data are extracted from massive volunteered geographic information OpenStreetMap databases through some data-intensive computing processes and four datasets on cities of **America (USA)**, **Britain** (UK), **France**, and **Germany** are formed. For simplicity, defining $r_f=2$, we can construct four self-similar hierarchies of the Euramerican cities. The values of the hierarchical scaling exponent (fractal dimension) and related parameter/statistics are tabulated as follows (Table A1).

**Table A1** The scaling exponents and related parameters/statistics of four self-similar hierarchies of Euramerican cities in 2010

| Parameter/Statistics | America (USA) | Britain (UK) | France | Germany |
|---|---|---|---|---|
| Fractal dimension ($D$) | 1.046 | 0.949 | 0.928 | 1.025 |
| Standard error ($\sigma$) | 0.008 | 0.039 | 0.039 | 0.020 |
| Goodness of fit ($R^2$) | 0.999 | 0.987 | 0.986 | 0.996 |
| City number ($n$) | 31305 | 1251 | 1240 | 5160 |
| Level number ($M$) | 15 | 11 | 11 | 13 |
| Scaling range (including levels) | $m=3\sim15$ | $m=1\sim10$ | $m=1\sim10$ | $m=2\sim12$ |

**Note**: The original city datasets of America (USA), Britain (UK), France, and Germany is available from: http://fromto.hig.se/~bjg/scalingdata/ .